\documentclass{article}

\begin{document}

\title{Positivity bounds on parton distributions at large~$N_{c}$}
\author{P.V. Pobylitsa}
\date{}
\maketitle

\begin{center}
\emph{Institute for Theoretical Physics II, Ruhr University Bochum,\\[0pt]
D44780, Bochum, Germany}\\[0pt]
and\\[0pt]
\emph{Petersburg Nuclear Physics Institute, 188350, Gatchina, Russia}
\end{center}

\begin{abstract}
Positivity bounds for twist-two parton distributions in
multicolored QCD are stronger than their analogs at finite $N_c$.
They include the enhanced large $N_{c}$ version of Soffer
inequality. These  bounds are compatible with the DGLAP evolution
to higher normalization points.
\end{abstract}

\section{Introduction}

The spectrum of the applications of the $1/N_{c}$ expansion \cite{Hooft-74}
to the analysis of the dynamics of strong interactions is rather rich.
However, the parton distributions of hadrons remained a sort of exception
for a long time. Historically the large $N_{c}$ behavior of parton
distributions was first analyzed in the context of the chiral quark soliton
model \cite{DPPPW-96,PP-1996} but the general structure of the $N_{c}
$ counting in this model is the same as in QCD.

Among recent applications of the large $N_{c}$ expansion to parton
distributions one should mention the phenomenological analysis of the
antiquark polarized distribution \cite{DGPW-99}, of the polarized gluon
distribution \cite{EGP-2000} and the large $N_{c}$ relations between the
quark distributions in the nucleon and in the $\Delta $ resonance
\cite{Chen-Ji-01}. Here we shall consider another interesting result:
combining the large $N_{c}$ expansion with the probabilistic interpretation
of parton distributions one can derive certain inequalities for parton
distributions \cite{PP-2000} which are \emph{stronger} than the
corresponding inequalities known in the case of finite $N_{c}$.

Our analysis will be based on the standard picture of the large $N_{c}$
baryons \cite{Witten-79} as states that can be described by mean field
(Hartree) equations of some effective theory. Although the exact action of
this effective theory equivalent to the large $N_{c}$ QCD is not known
certain conclusions can be made using only the assumption about the
symmetry \cite{Witten-83, Balachandran-83,Dashen-94} of the
solutions of the corresponding Hartree equations.

The large $N_{c}$ inequalities described here are not mere products of the
large $N_{c}$ pure art per se. For example, they include the enhanced large $
N_{c}$ version of Soffer inequality \cite{Soffer-95} for the transversity
distribution that can be of interest for the phenomenological applications.
Actually some of the results considered here were noticed earlier \cite
{PP-1996} in the context of the quark-soliton model \cite{DPPPW-96}. Since
the derivation was based only on the spin-flavor symmetry of the large $N_{c}
$ baryons the generalization to the case of the large $N_{c}$ QCD is rather
straightforward.

\section{Positivity bounds at finite $N_{c}$}

Let us first recall the well-known positivity properties of parton
distributions in the real world at finite number of colors. The unpolarized
parton distribution $q_{f}(x)$ is defined as the probability to find a quark
of type $f$ carrying fraction $x$ of the nucleon momentum in the
infinite momentum frame. The positivity of this probability leads us to the
simplest inequality $q_{f}(x)\geq 0$. If one considers the longitudinally
polarized nucleon then $q_{f}(x)$ can be decomposed into two pieces coming
from quarks with the longitudinal spin collinear with nucleon momentum ($
q_{f}^{\uparrow }(x)$) and with anticollinear spin ($q_{f}^{\downarrow }(x)$)
\begin{equation}
q_{f}(x)=q_{f}^{\uparrow }(x)+q_{f}^{\downarrow }(x)\,.  \label{q-decompos}
\end{equation}
Since both functions $q_{f}^{\uparrow }(x)$ and $q_{f}^{\downarrow }(x)$
have a probabilistic interpretation they are positive
\begin{equation}
q_{f}^{\uparrow }(x)\geq 0\,,\quad q_{f}^{\downarrow }(x)\geq 0\,.
\label{q-pol-pos}
\end{equation}
The (longitudinally) polarized parton distribution is defined as
\begin{equation}
\Delta _{L}q_{f}(x)=q_{f}^{\uparrow }(x)-q_{f}^{\downarrow }(x)\,.
\label{q-pol-def}
\end{equation}
Now it follows from (\ref{q-decompos}), (\ref{q-pol-pos}), (\ref{q-pol-def})
\begin{equation}
|\Delta _{L}q_{f}(x)|\leq q_{f}\,.
\end{equation}
Next, one can consider the nucleon and quarks with the transverse
polarization and define by analogy with $\Delta _{L}q_{f}$ the transversity
distribution $\Delta _{T}q_{f}$ which will obey inequality
\begin{equation}
|\Delta _{T}q_{f}(x)|\leq q_{f}\,.
\end{equation}
One more inequality can be obtained by choosing arbitrary directions for the
polarizations of quarks and nucleons. The positivity of the corresponding
probabilities leads to Soffer inequality \cite{Soffer-95}
\begin{equation}
q_{f}+\Delta _{L}q_{f}\geq 2|\Delta _{T}q_{f}|\,.
\end{equation}

\section{Positivity bounds at large $N_{c}$}

Thus in the real world with $N_{c}=3$ (and at any other finite $N_{c}$) we
have the following set of the positivity bounds for the unpolarized
($q_{f}$), longitudinally polarized ($\Delta _{L}q_{f}$) and transversity ($\Delta
_{T}q_{f}$) quark distributions
\begin{equation}
\begin{array}{c}
\displaystyle|\Delta _{L}q_{f}|\leq q_{f}\,, \\
\displaystyle|\Delta _{T}q_{f}|\leq q_{f}, \\
\displaystyle2|\Delta _{T}q_{f}|\leq q_{f}+\Delta _{L}q_{f}\,.
\end{array}
\quad \quad {\rm (finite\ }N_{c}{\rm )}  \label{finite-N-c-ineq}
\end{equation}

Below we shall derive the following analogs of these inequalities in the
large $N_{c}$ limit:
\begin{equation}
\begin{array}{l}
\displaystyle
\vphantom{\int\limits_a^b}
\left| \Delta _{L}q_{f}\right| \leq \frac{1}{3}q_{f}\left[
1+O(N_{c}^{-1})\right] \,, \\
\displaystyle|\Delta _{T}q_{f}|\leq \frac{1}{2}q_{f}\left[ 1+O(N_{c}^{-1})
\right] \,, \\
\displaystyle\vphantom{\int\limits_a^b}
2|\Delta _{T}q_{f}|\leq \left( \frac{1}{3}q_{f}+\Delta
_{L}q_{f}\right) \left[ 1+O(N_{c}^{-1})\right] \,.
\end{array}
\quad \quad (N_{c}\rightarrow \infty )  \label{new-ineq}
\end{equation}
Note that these inequalities for the large $N_{c}$ case are \emph{stronger}
than the corresponding inequalities at finite $N_{c}$ (\ref{finite-N-c-ineq}) because of the factors $1/3$ and $1/2$ appearing in the right-hand sides.
The origin of these factors can be traced to the spin-flavor symmetry of the
large $N_{c}$ baryons; in particular, the factor of $1/3$ has nothing to do
with the fact that $1/N_{c}=1/3$ in the real world.

This enhancement of the positivity bounds in the large $N_{c}$ limit is
rather a striking result. Naively one could expect that in the leading order
of the
large $N_{c}$ limit some part of information about parton distributions
would be lost but instead in this limit we have new stronger bounds. The
reason of this phenomenon can be explained in simple terms without appealing to
the technical derivation of (\ref{new-ineq}). In the leading order of the
large $N_{c}$ approximation the nucleon is degenerate with the $\Delta $
resonance and other baryonic resonances with $T=J=1/2,3/2,5/2,\ldots $ For
all of these particles one can analyze the twist-two parton distributions
and derive the corresponding positivity bounds. Moreover, since all these
particles are degenerate at the level of the leading order of the $1/N_{c}$
expansion one can consider the formal superposition of states with different
values of $T=J$ and derive the positivity bounds for the parton distribution
function of these formal superpositions. Although the superpositions of
states with different $T=J$ make no physical sense this formal trick allows
us to derive the new enhanced positivity bounds (\ref{new-ineq}) valid in
the large $N_{c}$ limit.

\section{Large $N_{c}$ counting for parton distributions}

In order to derive the large $N_{c}$ inequalities (\ref{new-ineq}) we should
start from the analysis of the large $N_{c}$ behavior of the parton
distributions. This analysis can be performed using standard large $N_{c}$
counting rules of QCD (see e.g.
\cite{DPPPW-96,EGP-2000,Chen-Ji-01}). On the other hand, it is
well known that the same large $N_{c}$ behavior of various quantities can be
also extracted from the nonrelativistic quark model, Skyrme model, quark
soliton model etc. Below we briefly describe the large $N_{c}$ behavior of
parton distributions appealing mainly to the naive quark model and using
sum rules for parton distributions.

Note the twist-two parton distributions $q_{f}(x),\,\Delta
_{L}q_{f}(x),\,\Delta _{T}q_{f}(x)$ depend on Bjorken variable $x$ so that
the analysis of the large $N_{c}$ behavior should involve a convention about
the behavior of $x$ in this limit. From the physical point of view the
nucleon consists of $N_{c}$ quarks (with an admixture of quark-antiquark
pairs) so
that it is natural to expect that at large $N_{c}$ limit the nucleon
momentum in the infinite momentum frame is more or less uniformly
distributed between $O(N_{c})$ quarks and antiquarks. Then in the large
$N_{c}$ limit it makes sense to concentrate on $x=O(1/N_{c})$. This physical
argument shows that that the large $N_{c}$ behavior of various parton
distributions should have the form
\begin{equation}
q_{f}(x)=N_{c}^{2}\left[ \phi _{f}(N_{c}x)+O(N_{c}^{-1})\right] ,
\end{equation}
\begin{equation}
\Delta _{L,T}q_{f}(x)=N_{c}^{2}\left[ \Delta _{L,T}\phi
_{f}(N_{c}x)+O(N_{c}^{-1})\right] \,.
\end{equation}
Functions $\phi _{f},\,\Delta _{L}\phi _{f},\,\Delta _{T}\phi _{f}$ are
stable in the large $N_{c}$ limit so that the $x$ dependence of the form
$\phi _{f}(N_{c}x)$ implies that $x\sim N_{c}^{-1}$ at large $x$. The factor
of $N_{c}^{2}$ ensures the consistency of the large $N_{c}$ behavior
for various sum rules for parton distributions. For example for the baryon
charge sum rule
\begin{equation}
\sum\limits_{f}\int\limits_{0}^{1}dx\left[ q_{f}(x)-\bar{q}_{f}(x)\right]
=\sum\limits_{f}\int\limits_{-1}^{1}dx\,q_{f}(x)=N_{c}
\label{baryon-number-sum-rule}
\end{equation}
we find in the large $N_{c}$ limit
\[
\int\limits_{-1}^{1}dx\,q_{f}(x)\stackrel{N_{c}\rightarrow \infty }{
\rightarrow }N_{c}^{2}\int\limits_{-1}^{1}dx\,\phi _{f}(N_{c}x),
\]
\begin{equation}
\stackrel{N_{c}\rightarrow \infty }{\rightarrow }N_{c}\int
\limits_{-N_{c}}^{N_{c}}dy\,\phi _{f}(y)\stackrel{N_{c}\rightarrow \infty }{
\rightarrow }N_{c}\int\limits_{-\infty }^{\infty }dy\,\phi _{f}(y)=O(N_{c})
\end{equation}
which really agrees with the sum rule (\ref{baryon-number-sum-rule}).

In the leading order of the large $N_{c}$ limit we have for the $u$ and $d$
distribution functions of the proton \cite{DPPPW-96}:
\[
q_{u}=q_{d}\left[ 1+O(N_{c}^{-1})\right] \,,\quad \Delta _{L}q_{u}=-\Delta
_{L}q_{d}\left[ 1+O(N_{c}^{-1})\right] \,,
\]
\begin{equation}
\quad \Delta _{T}q_{u}=-\Delta _{T}q_{d}\left[ 1+O(N_{c}^{-1})\right]\,.
\label{q-u-q-d}
\end{equation}
These large $N_{c}$ relations have a simple interpretation in the
nonrelativistic quark model: at large odd $N_{c}$ the proton consists of
$(N_{c}+1)/2$ $u$-quarks and $(N_{c}-1)/2$ $d$-quarks. In the leading order
of large $N_{c}$ we have the same amount of $u$ and $d$ quarks which means
that $q_{u}=q_{d}\left[ 1+O(1/N_{c})\right] $. Next, since we want to have
the spin of the proton $1/2$, at large $N_{c}$ there must be a cancelation
between the $O(N_{c})$ parts of the spin of $u$ and $d$ quarks, which is
expressed by the last two equations (\ref{q-u-q-d}).

\section{Derivation of large $N_{c}$ inequalities}

\label{step-1-subsection}

Below we sketch the main ideas of the derivation of inequalities
(\ref{new-ineq}). The complete description can be found in ref. \cite{PP-2000}.

Although we cannot compute parton distributions in the large $N_{c}$ QCD,
still it is possible to write certain general relations using only the
spin-flavor symmetry of the large $N_{c}$ baryons. The twist-two quark
distributions $q_{u},\Delta _{L}q_{u},\Delta _{T}q_{u}$ in proton can be
represented in the form
\begin{eqnarray}
q_{u}(x) &=&\frac{1}{2}\mathrm{Sp}\rho (x)\,,  \nonumber \\
\Delta _{L}q_{u}(x) &=&-\frac{1}{6}\mathrm{Sp}\left[ \gamma ^{5}\tau
^{3}\rho (x)\right] ,  \label{q-rho} \\
\Delta _{T}q_{u}(x) &=&\frac{1}{12}\mathrm{Sp}\left[ \gamma ^{5}\left(
\gamma ^{1}\tau ^{1}+\gamma ^{2}\tau ^{2}\right) \rho (x)\right] \,.  \nonumber
\end{eqnarray}
Here $\rho _{f^{\prime }s^{\prime },fs}(x)$ is an $8\times 8$ matrix with
$SU(2)$ flavor indices $f^{\prime },f$ and with Dirac spin indices $s^{\prime
},s$. Matrix $\rho (x)$ depending on $x$ is determined by the dynamics of
the large $N_{c}$ QCD and is not known. Nevertheless we are aware of
the following general properties:

1) $\rho $ is a hermitean positive matrix (i.e. all eigenvalues are positive
or zero)
\begin{equation}
\rho ^{+}=\rho \,,
\end{equation}
\begin{equation}
\rho \geq 0\,,  \label{rho-positive-0}
\end{equation}

2) $\rho $ lives in the subspace of the projector $(1+\gamma ^{0}\gamma
^{3})/2$
\begin{equation}
\rho (\gamma ^{0}\gamma ^{3})=(\gamma ^{0}\gamma ^{3})\rho =\rho \,,
\end{equation}

3) $\rho $ commutes with $i\gamma ^{1}\gamma ^{2}+\tau ^{3}\equiv i\gamma
^{1}\gamma ^{2}\otimes 1_{flavor}+1_{spin}\otimes \tau _{3}$
\begin{equation}
\lbrack \rho ,(i\gamma ^{1}\gamma ^{2}+\tau ^{3})]=0\,.
\end{equation}

These properties allow us to write the following general representation
for $\rho (x)$

\begin{equation}
\rho (x)=\frac{1+\gamma ^{0}\gamma ^{3}}{2}\left[ c_{1}(x)\cdot \mathbf{1}
+c_{2}(x)\gamma ^{5}\tau ^{3}+c_{3}(x)\gamma ^{5}(\gamma ^{1}\tau
^{1}+\gamma ^{2}\tau ^{2})\right]  \label{rho-c}
\end{equation}
where $c_{i}(x)$ are some scalar functions of $x$.

The positivity property (\ref{rho-positive-0}) leads to the following
constraints on coefficients $c_{i}$
\begin{equation}
c_{1}-c_{2}\geq 2|c_{3}|,  \label{c-ineq-1}
\end{equation}
\begin{equation}
c_{1}+c_{2}\geq 0\,.  \label{c-ineq-2}
\end{equation}

Inserting the explicit representation for $\rho $ (\ref{rho-c}) into the
general equations for parton distributions (\ref{q-rho}) we immediately
express the parton distributions $f_{i}(x)$ in terms the coefficients
$c_{i}(x)$. Then the constraints (\ref{c-ineq-1}), (\ref{c-ineq-2}) on these
coefficients immediately lead to bounds (\ref{new-ineq}) on parton
distributions.

\section{Phenomenological status}

Concerning the practical applications of the large $N_{c}$ inequalities we
have to keep in mind two competing factors:

1) large $N_{c}$ inequalities (\ref{new-ineq}) are stronger than the
standard inequalities (\ref{finite-N-c-ineq}) valid at finite
$N_{c}$,

2) at finite $N_{c}$ inequalities (\ref{new-ineq}) can be violated by large
$1/N_{c}$ corrections.

In the case of $\Delta _{L}q_{f}$ the $1/N_{c}$ corrections are known to be
rather large. Indeed, the Bjorken sum rule relates $\Delta _{L}q_{u}-\Delta
_{L}q_{d}$ with the axial constant $g_{A}$ and it is well known that $g_{A}$
is usually underestimated in various model calculations based on the
leading order of the
$1/N_{c} $ expansion, which is usually attributed to sizable $1/N_{c}$
corrections. Hence the  $1/N_c$ corrections to $\Delta _{L}q_{f}$ should
be also large.
 Therefore one should not wonder that our large $N_{c}$
inequality for $\Delta _{L}q_{u}-\Delta _{L}q_{d}$
\begin{equation}
\frac{\left| \Delta _{L}q_{u}-\Delta _{L}q_{d}\right| }{q_{u}+q_{d}}\leq
\frac{1}{3}\left[ 1+O(N_{c}^{-1})\right]
\end{equation}
following from (\ref{new-ineq}), (\ref{q-u-q-d}) is in conflict with
experimental data. For example, in the GRSV parametrization \cite{GRSV-2001}
\begin{equation}
\max_{x}\frac{\left| \Delta _{L}q_{u}-\Delta _{L}q_{d}\right| }{q_{u}+q_{d}}
\sim 0.6\,.
\end{equation}

On the other hand, the large $N_{c}$ bounds (\ref{finite-N-c-ineq}) can be
of use in the case of the transversity distribution
and polarized antiquark distribution
where the experimental knowledge is rather poor.

\section{Concluding remarks}

In the large $N_{c}$ limit one has the positivity bounds
(\ref{finite-N-c-ineq}) for twist-two parton distributions that are stronger than
the standard positivity bounds at finite $N_{c}$. However, the practical
applications of the large $N_{c}$ inequalities are sensitive to the size of
the $1/N_{c}$ corrections.

In the case of finite $N_{c}$ it is well known that positivity bounds
(\ref{finite-N-c-ineq}) are compatible with the DGLAP evolution
equations in the following
sense. If inequalities (\ref{finite-N-c-ineq}) hold at some
normalization point then the one-loop evolution to higher normalization
points will not violate these inequalities. In ref. \cite{PP-2000} it was
shown that the large $N_{c}$ inequalities (\ref{new-ineq}) have the same
property: if one uses the evolution kernels only in the leading order of the
large $N_{c}$ expansion then inequalities (\ref{new-ineq}) are preserved by
the evolution upwards.

{\bf Acknowledgements.}

The results described here are obtained in collaboration with M.V. Polyakov.
I am thankful to A.~Belitsky, D.I.~Diakonov, A.V.~Efremov,
L.~Frankfurt, K.~Goeke, N.~Kivel, V.Yu.~Petrov, A.V.~Radyushkin, 
A.~Shuvaev, M.~Strikman and O. Teryaev
for the discussions. It is a pleasure to thank R. Lebed for the
organization of the workshop. The author appreciates the support of INT,
DFG and BMBF.

\end{document}